\documentclass[conference, 10pt]{IEEEtran}

\usepackage{amsmath,amssymb,amsfonts}
\usepackage{cite}
\usepackage{url}
\usepackage{graphicx}
\usepackage{color}
\usepackage{placeins}
\usepackage{float}
\usepackage{tabularx,colortbl}
\usepackage{ifthen}
\makeatletter

\newcounter{author}
\renewcommand{\author}[2][]{
   \stepcounter{author}
   \@namedef{author@\theauthor}{#2}
   \@namedef{authorlabel@\theauthor}{#1}
}

\newcounter{address}
\newcommand{\address}[2][]{
   \stepcounter{address}
   \@namedef{address@\theaddress}{#2}
   \@namedef{addresslabel@\theaddress}{#1}
}

\newcommand{\alsep}{and}

\def\newmaketitle{\par%
  \begingroup%
  \normalfont%
  \def\thefootnote{}%  the \thanks{} mark type is empty
  \def\footnotemark{}% and kill space from \thanks within author
  \let\@makefnmark\relax% V1.7, must *really* kill footnotemark to remove all \textsuperscript spacing as well.
  \footnotesize%       equal spacing between thanks lines
  \footnotesep 0.7\baselineskip%see global setting of \footnotesep for more info
  \normalsize%
  \twocolumn[\thenewmaketitle\@IEEEaftertitletext]%
  \if@IEEEusingpubid
     \enlargethispage{-\@IEEEpubidpullup}%
  \fi
  \endgroup
  \setcounter{footnote}{0}\let\maketitle\relax\let\@maketitle\relax
  \gdef\@thanks{}%
  \let\thanks\relax}

\def\thenewmaketitle{
  \newpage
  \begin{center}%
    \vskip0.2em{\Huge\@IEEEcompsoconly{\sffamily}\@IEEEcompsocconfonly{\normalfont\normalsize\vskip 2\@IEEEnormalsizeunitybaselineskip
   \bfseries\large}\@title\par}\vskip1.0em\par%
    \vspace{1ex}
    \newcounter{c@author}
    \newcounter{c@tmp}
    \ifthenelse{\value{author}=2}{%
      \newcommand{\liand}{ and }}{%
      \newcommand{\liand}{, and }}
    \ifthenelse{\value{address}<2}{%
      \@nameuse{author@1}%
      \stepcounter{c@author}%
      \whiledo{\value{c@author}<\value{author}}{%
        \setcounter{c@tmp}{\value{author}}%
        \addtocounter{c@tmp}{-\value{c@author}}%
        \ifthenelse{\value{c@tmp}=1}{%
          \renewcommand{\alsep}{\liand}}{\renewcommand{\alsep}{, }}%
        \stepcounter{c@author}\alsep \@nameuse{author@\thec@author}}\\%
    }
    {%Add address references after the author's name
      \@nameuse{author@1}${}^{(\ref{\@nameuse{authorlabel@1}})}$%
      \stepcounter{c@author}%
      \whiledo{\value{c@author}<\value{author}}{%
      \setcounter{c@tmp}{\value{author}}%
      \addtocounter{c@tmp}{-\value{c@author}}%
      \ifthenelse{\value{c@tmp}=1}{%
        \renewcommand{\alsep}{\liand}}{\renewcommand{\alsep}{, }}%
      \stepcounter{c@author}\alsep \@nameuse{author@\thec@author}%
        ${}^{(\ref{\@nameuse{authorlabel@\thec@author}})}$%
      }
    }
    \vspace{0.2ex}

    \ifthenelse{\value{address}>0}{%
      \ifthenelse{\value{address}=1}{
        {\@nameuse{address@1}}
      }
      {%Output the addresses as an enumerated list
        \newcounter{c@address}

        \begin{center}
        \whiledo{\value{c@address}<\value{address}}
        {
          \refstepcounter{c@address}
            ${}^{(\thec@address)}$\,%
              \label{\@nameuse{addresslabel@\thec@address}}%
              \@nameuse{address@\thec@address}\\ %
        }
        \end{center}
      } % end of the address creation ifthenelse block
    }
    {
      \relax
    }
  \end{center}
}

\makeatother

\title{Nonlocal Dual-Band Reconfigurable Intelligent Surfaces for Precise Full-Space Beamforming}

\author[org1]{Moosung Kim}
\author[org1]{Minseok Kim}

\address{School of Electronic and Electrical Engineering, Hongik University, 94 Wausan-ro, Mapo-gu, Seoul 121-791, Korea,      minseok.kim@hongik.ac.kr}

\begin{document}

\newmaketitle

\begin{abstract}
This paper introduces a nonlocal, dual-band reconfigurable intelligent surface (RIS) designed for full-space beam synthesis at 4.0 GHz and 6.3 GHz. The constituent unit cells comprise a pair of interleaved sub-cells that are specifically engineered to operate independently at their respective target frequencies. This hardware-level decoupling facilitates an efficient synthesis framework based on microwave network theory (MNT) that rigorously accounts for mutual coupling within both bands. Under this framework, the optimal biasing for sub-cells is determined to achieve precise full-space beam synthesis at both frequencies. The proposed method is numerically and experimentally validated with an RIS comprising $14 \times 14$ varactor-loaded unit cells that can be individually biased. We experimentally demonstrate arbitrary beam profile synthesis beyond simple beam steering, including dual-beam and sector patterns in full space. Experimental and simulation results show good agreement with the MNT model, confirming the effectiveness of the proposed method.%괜찮아

\end{abstract}

%. To be effective, such dual-band systems must provide simultaneous services at distinct frequencies while ensuring high spectral isolation and independent wavefront control

\section{Introduction}
Reconfigurable intelligent surfaces (RISs) have emerged as a key enabling technology for beyond-5G/6G systems, providing the capability to reconfigure wireless channels by dynamically controlling the scattered electromagnetic (EM) waves. This ability to tailor the EM environment allows for the mitigation of fundamental propagation hurdles, such as signal blockages and multipath fading~\cite{Renzo2020JSAC}. 
%flow1: RIS의 일반적인 필요성,RIS에 대한 간단한 설명.
Nevertheless, as these wireless environments grow increasingly complex, the functional requirements for RISs are evolving toward arbitrary beamforming beyond simple point-to-point beam steering. Moreover, the growing demand for multifunctional links, such as integrated sensing and communication (ISAC), makes dual-band operation a critical requirement~\cite{Chen2025TAP}.
%flow2: 통신기술의 발달에따른 RIS에 대하여 요구되어지는 스펙증가(arbi-beamshaping,dual_band). 

Despite these demands, conventional RISs often fall short as they typically rely on the local periodicity assumption. Indeed, RIS unit cells are commonly modeled under periodic boundary conditions within commercial full-wave simulators, such as \texttt{ANSYS HFSS}, which implicitly assumes that the physical properties of neighboring unit cells remain uniform across the surface. However, as functional RISs feature spatially varying surface profiles to synthesize specific beams, this local periodicity is merely an approximation. Consequently, complex mutual coupling and nonlocal interactions between unit cells are overlooked, leading to significant discrepancies and the eventual failure of accurately synthesizing desired far-field patterns.
 %flow3: 이러한 요구조건(arbi-beeamshaping)을 만족시키지 못하는 기존의 한계(LPC의 채택으로 인해).
 
To overcome the limitations of conventional local RISs, several nonlocal design methodologies have been proposed. One approach utilizes integral equation-based formulations to model the entire surface, enabling global optimization of the scattering response via the method of moments~\cite{Budhu2021IEEETAP, Kim2025PRApplied}. While rigorous, the computational complexity of these methods increases exponentially for large-scale, two-dimensional (2D) reconfigurable surfaces, particularly when dual-band optimization is required.
%flow4: 이러한 요구조건(arbi-beeamshaping)을 만족시키기 위해 LPC대신 IE를 채택하여 일부 극복했으나 여전히 다른 요구조건(dual-band),(active),(fullspace shaping) 은 만족시키지 못함.
On the other hand, MNT-based methods model the RIS as a multiport network with tunable loads. This approach significantly reduces computational complexity, offering a more efficient alternative for large-scale surfaces~\cite{Almunif2025TAP, Qiu2025access}. Nevertheless, extending such a network-based framework to dual-band RISs still remains challenging, as cross-band interference often complicates the synthesis within a unified analytical model.
%flow5: IE의 단점인 computational cost를 어느정도 극복한 MNT 방법 제시. 하지만 이방법도 여전히 band간 간섭까지 고려하기 어려워 모든 요구 조건을 만족시키기 힘들다.(dual-band operation)

To address these challenges, we present a dual-band nonlocal RIS designed through an MNT-based framework that leverages a unit-cell architecture with interleaved sub-cells. Specifically, the unit cell is composed of two distinct sub-elements, each specifically resonant at 4.0 GHz and 6.3 GHz, which ensures independent control over each frequency. This hardware-level decoupling allows the network representation to be independently formulated for each band, enabling the simultaneous nonlocal synthesis of arbitrary beam profiles across both frequencies with minimal cross-band interference. The versatility of the proposed approach is numerically and experimentally validated through an RIS comprising $14 \times 14$ varactor diode-loaded unit cells. In particular, various complex beamforming capabilities in full space, including multi-beam and sector patterns, are demonstrated, which show excellent agreement with the analytical predictions of the MNT model.
%flow6: 요구조건이 모두 만족시키기 위해 hardware level에서 우선 band간 간섭을 저지, MNT 방법 적용하는 아키텍쳐 제안. 시뮬,실험을 통해 이것이 검증 되었다.
\section{Proposed Unit-Cell Architecture}
\begin{figure}[t!]
  \centering
  \includegraphics[height=100mm]{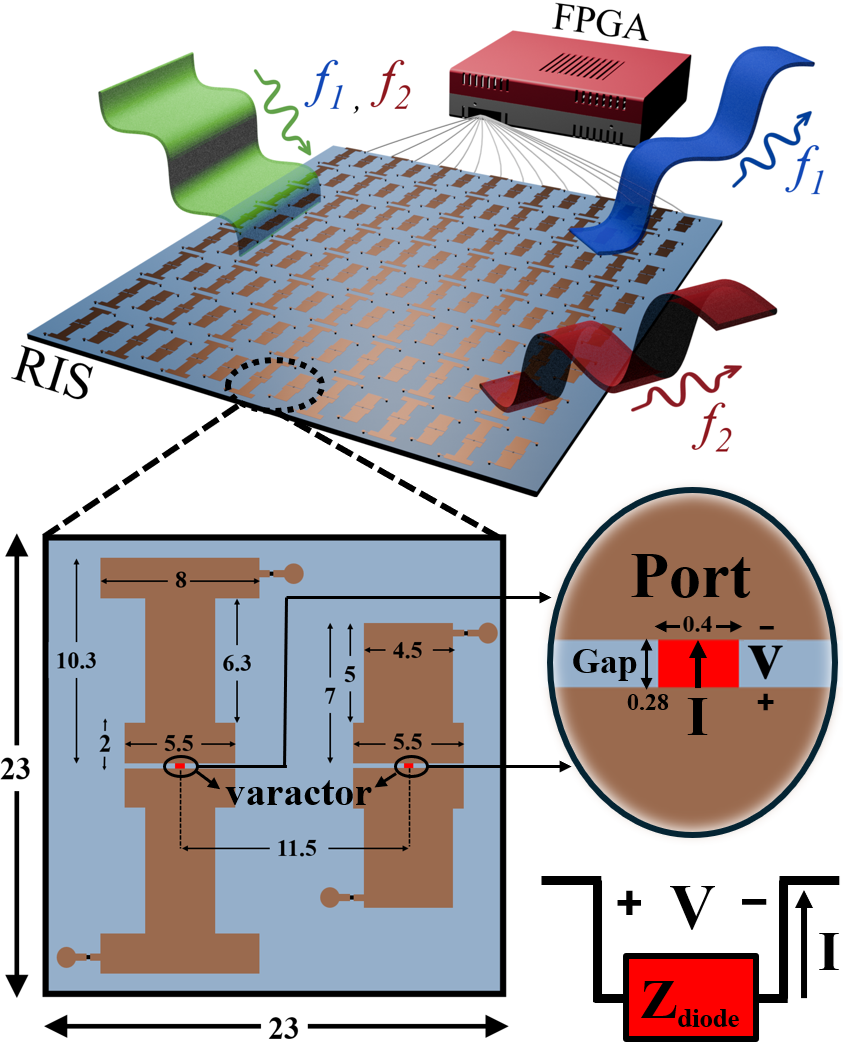}
  \caption{The proposed band-decoupled dual-band reconfigurable intelligent surface. The unit cell’s
  dimensions are shown in millimeters (mm).}
  \label{fig1}
\end{figure}

Fig.~\ref{fig1} illustrates the proposed dual-band reconfigurable unit cell, which comprises two interleaved sub-cells tailored for $\hat{x}$-polarized waves at $f_1 = 4.0$ GHz and $f_2 = 6.3$ GHz.
% \section{Proposed Unit-Cell Architecture}
% Fig.~\ref{fig1} illustrates the proposed dual-band reconfigurable unit cell, which comprises two interleaved sub-unit cells tailored for $\hat{x}$-polarized waves at $f_1 = 4.0$ GHz and $f_2 = 6.3$ GHz.
% %
% \begin{figure}[t!]
% 	\centering
% 	\resizebox{0.45\textwidth}{!}{\includegraphics{2026_URSI_Figures/Network_representation.png}}
% 	\caption{The proposed band-decoupled dual-band reconfigurable intelligence surface. The unit cell’s
% dimensions are shown in millimeters (mm).}
% 	\label{fignr}
% \end{figure}
% %
For clarity, the sub-cells operating at $f_1$ and $f_2$ are henceforth denoted as Cell-L and Cell-U, respectively. The unit cell is integrated onto a grounded TMM-4 substrate ($\varepsilon_r = 4.7$, $\tan\delta = 0.002$) with a thickness of 1.9 mm. To ensure subwavelength sampling of surface currents for precise wavefront control, Cell-L and Cell-U are designed with compact electrical sizes of $0.307\lambda_1$ and $0.483\lambda_2$ at their respective operating frequencies. Each sub-cell incorporates a varactor diode (MACOM MAVR-000120), providing a continuous capacitance range from 0.2 pF to 1.0 pF to enable full reconfigurability.

\begin{figure}[t!]
	\centering
	\resizebox{0.45\textwidth}{!}{\includegraphics{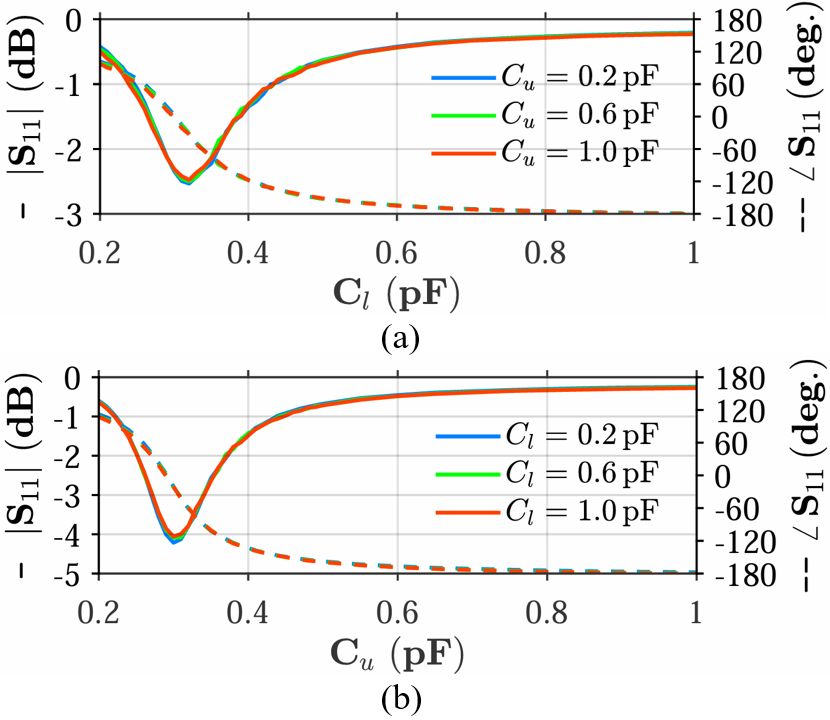}}
	\caption{The reflection coefficient of the unit cell at (a) $f_1$ and (b) at $f_2$ under periodic boundary conditions, illustrating that the modulation of sub-cells at $f_1$ (or $f_2$) is transparent at $f_2$ (or $f_1$).}
	\label{fig2}
\end{figure}

A primary design objective here is to ensure that the biasing of one sub-cell does not perturb the EM response of the other, thereby achieving hardware-level isolation. To this end, the sub-cells are strategically interleaved such that their dominant current paths are spatially separated, thereby minimizing cross-band interference. This decoupling is verified in Fig.~\ref{fig2}, which shows the reflection responses under periodic boundary conditions for $\hat{x}$-polarized illumination. It is noted that these responses are evaluated at an oblique incidence of $15^\circ$ to accurately represent the experimental conditions, where the fabricated RIS is excited by a feed horn positioned $15^\circ$ off-broadside.

In Fig.~\ref{fig2}, $C_{l}$ and $C_{u}$ denote the varactor capacitances loaded on Cell-L and Cell-U, respectively. As shown in Fig.~\ref{fig2}(a), the reflection response at $f_1$ is primarily governed by Cell-L; varying $C_{l}$ yields a reflection phase tuning range of $290^\circ$ with a maximum loss of 2.5 dB at $f_1$, while the influence of $C_{u}$ remains negligible. Similarly, Fig.~\ref{fig2}(b) demonstrates that the response at $f_2$ is predominantly controlled by Cell-U. In particular, a phase tuning range of nearly $300^\circ$ is achieved, with a corresponding loss of 4.1 dB at $f_2$, whereas varying $C_{l}$ produces minimal impact. These results confirm that, under periodic boundary conditions, the proposed unit cell exhibits highly decoupled operation across the two target frequencies.

Building upon this observation, we adopt the pragmatic assumption—later validated experimentally in Section V—that this band-decoupled behavior remains robust even within inhomogeneous environments. This allows for a simplified yet rigorous extension of the MNT-based design to dual-band operation. By leveraging this frequency-domain isolation, the following section details a nonlocal synthesis framework that enables independent and precise wavefront control for the proposed RIS.

\section{MNT-Based Nonlocal Design Framework}
With the unit-cell geometries defined, the MNT-based approach is employed to synthesize the desired far-field patterns \textit{independently} at $f_1$ and $f_2$. This method first models the RIS without any varactor diodes at each frequency as an impedance matrix, $\mathbf{Z}^\text{nf}$, which characterizes both the self-impedance and mutual coupling between constituent sub-cells. Within this framework, the surface currents $\mathbf{I}$ and the voltages $\mathbf{V}$ across the designated gaps where the varactors are to be placed (see Fig. \ref{fig1}) are linked as,
\begin{equation}
\mathbf{V} = \mathbf{Z}^\text{nf} \mathbf{I} + \mathbf{V}_{\mathrm{inc}} ,
\label{eq:VmI}
\end{equation}
where $\mathbf{V}_{\mathrm{inc}}$ denotes the voltage induced across the gaps by the incident field.

%This formulation abstracts each diode location as a virtual port to numerically construct $\mathbf{Z}^\text{nf}$ and $\mathbf{V}_{\mathrm{inc}}$ through full-wave simulations in \texttt{ANSYS HFSS}. 

This formulation models each potential diode site as a virtual port, allowing us to extract $\mathbf{Z}^\text{nf}$ and $\mathbf{V}_{\mathrm{inc}}$ through full-wave simulations in \texttt{ANSYS HFSS}. Specifically, the impedance element $Z^\text{nf}_{nm}$ is extracted by impressing a unit line current (1 A) at the $m^\text{th}$ diode location and integrating the resulting electric field across the $n^\text{th}$ diode's gap to determine the induced voltage. The impedance value is then defined as the ratio of this voltage to the excitation current. Similarly, the elements of $\mathbf{V}_{\mathrm{inc}}$ are obtained by illuminating the diode-free RIS with the target incident wave and recording the induced voltage across each port using the same integration method. This procedure is carried out for an RIS comprising $14 \times 14$ unit cells, which results in $\mathbf{V}, \mathbf{I}, \mathbf{V}_{\mathrm{inc}} \in \mathbb{C}^{14 \times 1}$ and $\mathbf{Z}^\text{nf} \in \mathbb{C}^{14 \times 14}$ matrices.

%In \eqref{eq:VmI}, $\mathbf{V}$ and $\mathbf{I}$ represent the voltages and currents at virtual ports defined across the designated gaps where the varactors are to be placed. This formulation allows us to determine the voltage induced across these gaps for any arbitrary current $\mathbf{I}$ impressed within the diode-free RIS structure. 

To account for the physical presence of the varactor diodes, the terminal behavior at these ports is additionally constrained by Ohm’s law as $\mathbf{V} = \mathbf{Z}_{\mathrm{d}} \mathbf{I}$, where $\mathbf{Z}_{\mathrm{d}} \in \mathbb{C}^{14 \times 14}$ is a diagonal matrix containing the load impedances of each diode. By substituting this into Eq.~\eqref{eq:VmI}, the port current distribution is expanded as
\begin{equation}
\mathbf{I} = \left( \mathbf{Z}_{\mathrm{d}} - \mathbf{Z}^\text{nf} \right)^{-1}
\mathbf{V}_{\mathrm{inc}} .
\label{eq:current}
\end{equation}
Since $\mathbf{Z}^\text{nf}$ and $\mathbf{V}_{\mathrm{inc}}$ are determined solely by the RIS geometry and the incident field, they remain constant for a fixed configuration. Consequently, the current distribution $\mathbf{I}$ becomes a function exclusively of the load impedance matrix $\mathbf{Z}_{\mathrm{d}}$.

By utilizing the elements of $\mathbf{Z}_{\mathrm{d}}$ as design variables, the current $\mathbf{I}$ is tailored to synthesize the desired far-field patterns. To this end, the total far field, $\mathbf{E}_{\mathrm{ff}}^{p}$, is directly related to the port currents through the superposition:
\begin{equation}
\mathbf{E}_{\mathrm{ff}}^{p}
= \mathbf{G}_{\mathrm{ff}}^{p}\mathbf{I}
+ \mathbf{E}_{\mathrm{fi}}^{p},
\label{eq:farfield}
\end{equation}
where $\mathbf{G}^{p}_{\mathrm{ff}}$ represents the Green’s function relating the port currents to the scattered far field, and $\mathbf{E}^{p}_{\mathrm{fi}}$ denotes the far-field contribution from the incident wave in the absence of port currents (i.e., the passive scattering of the diode-free RIS), with the superscript $p \in \{\theta,\phi\}$ indicating the polarization state of the resulting field.
 Similar to the previous extraction process, $\mathbf{G}_{\mathrm{ff}}^{p}$ is obtained by impressing a unit line current at the $m^\text{th}$ port and recording the resulting far-field response across the radiation boundary in \texttt{ANSYS HFSS} as a function of $\theta$ and $\phi$.

Since \eqref{eq:current} and \eqref{eq:farfield} hold independently at the two operating frequencies under the proposed unit cell, the following relation holds:
\begin{equation}
\mathbf{E}_{\mathrm{ff}}^{p,(k)}
=
\mathbf{G}_{\mathrm{ff}}^{p,(k)}
\left(
\mathbf{Z}_{\mathrm{d}}^{(k)} - \mathbf{Z}^{\text{nf},(k)}
\right)^{-1}
\mathbf{V}_{\mathrm{inc}}^{(k)}
+
\mathbf{E}_{\mathrm{fi}}^{p,(k)},  k \in \{1,2\}.
\label{eq:Eff_Zd}
\end{equation}
where the superscripts $k=1$ and $k=2$ correspond to $f_1$ and $f_2$, respectively. In \eqref{eq:Eff_Zd}, the only design variables are the port load impedances $\mathbf{Z}_{\mathrm{d}}^{(1,2)}$. By optimizing these impedances, the desired far-field responses $\mathbf{E}_{\mathrm{ff}}^{p,(1,2)}$ are synthesized independently at each frequency band.

\section{Optimization of Varactor Capacitance}
The optimization is carried out to determine the port load impedances $\mathbf{Z}_{\mathrm{d}}^{(1,2)}$ by utilizing a genetic algorithm (GA). Accordingly, the cost function is defined as
\begin{equation}
\label{eq:multibeam_cost}
\begin{aligned}
f_1 &=
- W_1 \sum_{k=1}^{N_b} D_k
+ W_2\,\mathrm{STD}\!\left(D_1, D_2, \ldots, D_{N_b}\right) \\
&\quad
+ W_3 \sum_{k=1}^{N_b}
\Big[
\big(\theta_{\mathrm{peak},k}-\theta_{\mathrm{des},k}\big)^2
+ \big(\phi_{\mathrm{peak},k}-\phi_{\mathrm{des},k}\big)^2
\Big] .
\end{aligned}
\end{equation}
Here, $N_b$ denotes the number of desired beams, and $D_k$ represents the peak directivity of the $k^\text{th}$ beam, ordered in descending magnitude. The term $\mathrm{STD}(\cdot)$ denotes the standard deviation of ${D_k}$ and is used to promote beam-to-beam uniformity. The pairs $(\theta_{\mathrm{peak},k}, \phi_{\mathrm{peak},k})$ and $(\theta_{\mathrm{des},k}, \phi_{\mathrm{des},k})$ correspond to the actual and target beam directions in spherical coordinates, respectively. The weighting factors $W_1$, $W_2$, and $W_3$ are used to balance peak directivity enhancement, beam-to-beam uniformity, and steering accuracy.
The proposed formulation directly synthesizes uniform multi-beam patterns and can be readily extended to sector-beam synthesis. Specifically, by densely placing the target directions ${(\theta_{\mathrm{des},k},\phi_{\mathrm{des},k})}_{k=1}^{N_b}$ within the desired angular region, the optimization naturally yields a sector-shaped radiation pattern.

\section{Full-wave and Experimental Verifications}
In this section, we numerically and experimentally validate the proposed design framework using an RIS composed of $14 \times 14$ unit cells. The prototype is fabricated by integrating two modular RISs comprising $14 \times 7$ unit cells, demonstrating the scalability of the design for larger aperture expansions. The experimental setup is illustrated in Fig.~3, where a horn antenna that is positioned $40$~cm from the RIS center illuminates the surface at an elevation angle of $-15^\circ$.
\begin{figure}[t!]
\centering
\includegraphics[width=0.97\columnwidth, height=0.8\columnwidth, keepaspectratio]{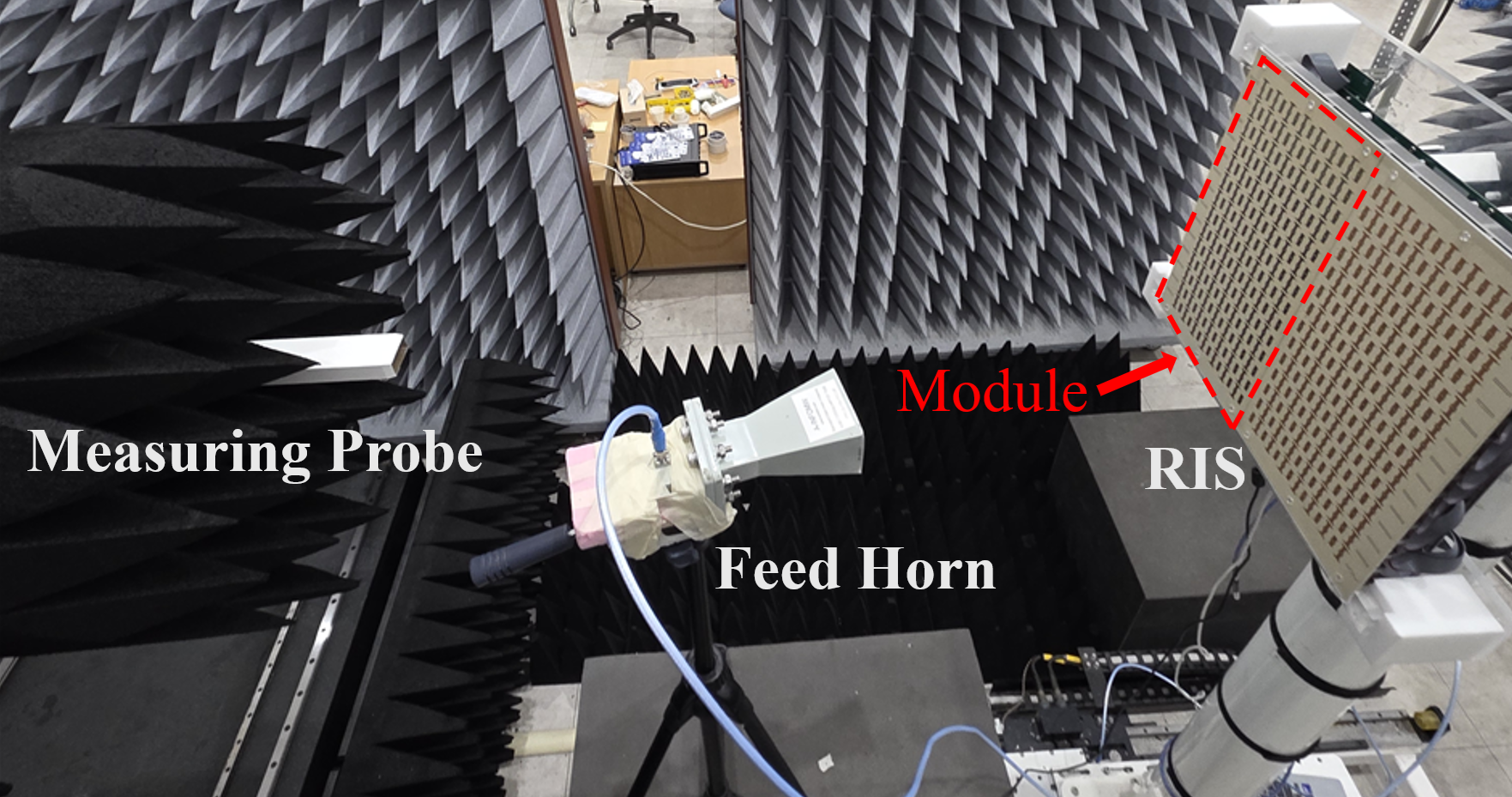}
\caption{Near-field measurement setup of the dual-band RIS: feed horn and a $14\times14$ RIS panel composed of two $14\times7$ modules.}
\label{fig3}
\end{figure}

\begin{figure*}[t!]
\centering
\includegraphics[width=\textwidth]{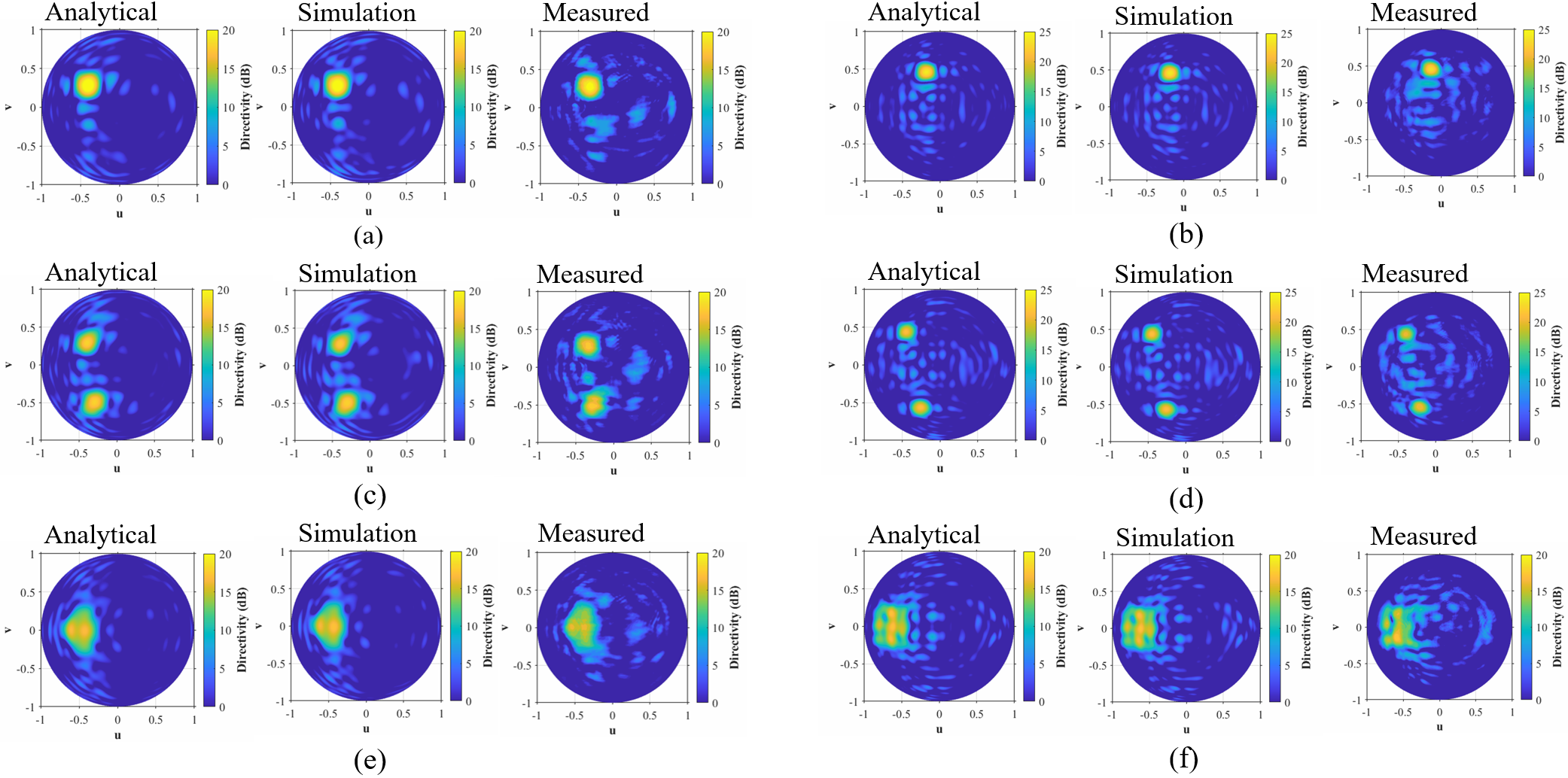}
\caption{UV-plane radiation patterns for the synthesized dual-band RIS: 
(a),(b) unidirectional beam steering at $f_1$ and $f_2$, respectively; 
(c),(d) dual-beam synthesis at $f_1$ and $f_2$, respectively; and 
(e),(f) sector-pattern synthesis at $f_1$ and $f_2$, respectively.}
% UV-plane radiation patterns for the synthesized dual-band RIS: (a),(b) MATLAB (MNT-based) predictions, (c),(d) HFSS full-wave simulations, and (e),(f) patterns reconstructed from near-field measurements. (a),(c),(e) correspond to $f_2$, demonstrating sector-beam synthesis, and (b),(d),(f) correspond to $f_1$, demonstrating dual-beam synthesis.
\label{fig4}
\end{figure*}

To demonstrate the versatility of the proposed framework, we evaluate three distinct scenarios: (i) single-beam steering, (ii) dual-beam generation, and (iii) sector pattern synthesis at $f_1$ and $f_2$. For each case, the optimization framework detailed in Sections III and IV is employed to determine the required load impedances, which are then physically realized by biasing the varactor diodes accordingly. Fig. \ref{fig4} summarizes the results, comparing the MNT-based analytical predictions, full-wave simulations, and measured near-field data.

Across all scenarios, the MNT predictions and full-wave simulations exhibit near-perfect agreement, confirming that the MNT formulation accurately characterizes beam patterns by rigorously accounting for mutual coupling. Although the operating frequencies of the fabricated RIS are shifted by $+100$~MHz due to fabrication tolerances, the measured patterns align well with the numerical results, experimentally validating the full-space beam synthesis capabilities of the proposed dual-band design. To the best of the authors' knowledge, this work represents the first experimental realization of sector pattern synthesis using a dual-band RIS in full space.

\begin{figure}[!t]
\centering
\includegraphics[width=1\columnwidth, height=1\columnwidth, keepaspectratio]{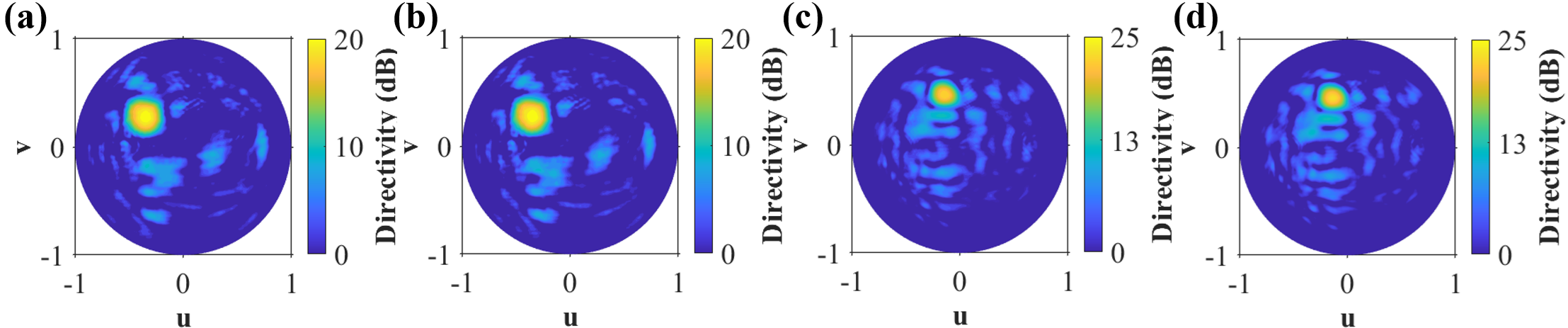}
\caption{Measured $u$--$v$ directivity maps demonstrating the band-decoupled characteristics of the dual-band RIS.}
\label{fig5}
\end{figure}

Notably, these results provide empirical evidence for the pragmatic assumption introduced in Section II: that the band-decoupled behavior remains robust even when unit cells are arranged in inhomogeneous environments. Fig. \ref{fig5} explicitly demonstrates this by illustrating the measured radiation patterns under various bias configurations.
In Fig. \ref{fig5}(a), the radiation pattern at $f_1$ is shown for an RIS biased to produce a single steered beam at both $f_1$ and $f_2$. In Fig. \ref{fig5}(b), the radiation pattern at $f_1$ is re-measured after the bias at $f_2$ is reconfigured to generate a dual-beam, while the bias at $f_1$ remains unchanged. The resulting pattern at $f_1$ is nearly identical to the previous case, indicating that the reconfiguration of bias at $f_2$ does not perturb the scattering response at $f_1$. The corresponding measurements at $f_2$—shown in Figs.~\ref{fig5}(c) and \ref{fig5}(d) exhibit the same behavior, where the response at $f_2$ remains unaffected by modifications to the bias states of the $f_1$ sub-cells. This stability at one band while the radiation pattern is reconfigured at the other band confirms that the band-decoupled behavior remains robust even when unit cells are arbitrarily biased. As such, these results provide experimental confirmation that the proposed RIS maintains strictly band-independent operation.

\section{Conclusion}
This paper presents a nonlocal, dual-band RIS designed for independent wavefront control at 4.0 GHz and 6.3 GHz. By leveraging a hardware-level decoupling strategy through interleaved sub-cells, the proposed architecture enables an MNT-based synthesis framework that rigorously accounts for mutual coupling while eliminating cross-band interference. Experimental validation using a $14 \times 14$ prototype confirms the versatility of the proposed framework by demonstrating various complex far-field syntheses, including dual-beam and sector patterns in full space. %This work provides a scalable and precise methodology for the multi-functional beam shaping essential for beyond-5G and 6G wireless infrastructures.

\section*{ACKNOWLEDGEMENT}
This work was supported by the National Research Foundation of Korea (NRF) grant funded by the Korea government(MSIT)(RS-2024-00341191).

% used to balance the columns on the last page adjust value as needed
%\IEEEtriggeratref{8}
% The "triggered" command can be changed if desired:
%\IEEEtriggercmd{\enlargethispage{-5in}}

% references section

% can use a bibliography generated by BibTeX as a .bbl file
% BibTeX documentation can be easily obtained at:
% http://www.ctan.org/tex-archive/biblio/bibtex/contrib/doc/
% The IEEEtran BibTeX style support page is at:
% http://www.michaelshell.org/tex/ieeetran/bibtex/
%\bibliographystyle{IEEEtran}
% argument is your BibTeX string definitions and bibliography database(s)
%\bibliography{IEEEabrv,../bib/paper}
%
% <OR> manually copy in the resultant .bbl file
% set second argument of \begin to the number of references
% (used to reserve space for the reference number labels box)

\bibliographystyle{IEEEtran}
\bibliography{MinseokKim_Refs}  

@ARTICLE{Almunif2025TAP,
  author={Almunif, Malik and Alsolamy, Faris and Grbic, Anthony},
  journal={IEEE Transactions on Antennas and Propagation}, 
  title={Network-Based Design of Reactive Beamforming Metasurfaces}, 
  year={2025},
  volume={73},
  number={8},
  pages={5970-5980},
  doi={10.1109/TAP.2025.3562760}}

@ARTICLE{Qiu2025access,
  author={Qiu, Tianke and Eleftheriades, George V.},
  journal={IEEE Access}, 
  title={A Design Framework for Reconfigurable Intelligent Metasurfaces Enabling Full-Space Beamforming Using Auxiliary Surface Waves}, 
  year={2025},
  volume={13},
  number={},
  pages={51969-51977},
  doi={10.1109/ACCESS.2025.3552755}}

@ARTICLE{Chen2025TAP,
  author={Chen, Kangjian and Qi, Chenhao and Dobre, Octavia A.},
  journal={IEEE Transactions on Communications}, 
  title={DBRAA: Sub-6 GHz and Millimeter Wave Dual-Band Reconfigurable Antenna Array for ISAC}, 
  year={2025},
  volume={73},
  number={10},
  pages={9830-9845},
  doi={10.1109/TCOMM.2025.3567007}}

@ARTICLE{Renzo2020JSAC,
  author={Di Renzo, Marco and Zappone, Alessio and Debbah, Merouane and Alouini, Mohamed-Slim and Yuen, Chau and de Rosny, Julien and Tretyakov, Sergei},
  journal={IEEE Journal on Selected Areas in Communications}, 
  title={Smart Radio Environments Empowered by Reconfigurable Intelligent Surfaces: How It Works, State of Research, and The Road Ahead}, 
  year={2020},
  volume={38},
  number={11},
  pages={2450-2525},
  doi={10.1109/JSAC.2020.3007211}}

@PREAMBLE{
 "\providecommand{\noopsort}[1]{}" 
 # "\providecommand{\singleletter}[1]{#1}%" 
}

@ARTICLE{Budhu2021IEEETAP,
  author={Budhu, Jordan and Grbic, Anthony},
  journal={IEEE Transactions on Antennas and Propagation}, 
  title={Perfectly Reflecting Metasurface Reflectarrays: Mutual Coupling Modeling Between Unique Elements Through Homogenization}, 
  year={2021},
  volume={69},
  number={1},
  pages={122-134},
  doi={10.1109/TAP.2020.3001450}}

@ARTICLE{Kim2025PRApplied,
  title = {Cavity-Excited Impedance Surfaces for Full-Space Beam Steering via Combination of Integral Equations and Baffles},
  volume = {23},
  number = {2},
  journal = {Physical Review Applied},
  author = {Kim, M.},
  year = {2025},
  month = feb,
  pages = {024011}
}
\end{document}